\documentclass{desyproc}

\newcommand{\Aprime}{\ensuremath{\mathrm{A}^\prime}}

\begin{document}

\title{The Heavy Photon Search Experiment}

\author{{\slshape Per Hansson Adrian$^1$}\\[1ex]
$^1$SLAC National Accelerator Laboratory, Menlo Park, CA, USA}

\contribID{familyname\_firstname}

\desyproc{DESY-PROC-2012-04}
\acronym{Patras 2012} 
\doi  

\maketitle


\begin{abstract}
Interest in new physics models including so-called hidden sectors has increased in 
recent years as a result of anomalies from astrophysical observations. The Heavy Photon Search (HPS)
experiment proposed at Jefferson Lab will look for a mediator of a new force, a 
GeV-scale massive U(1) vector boson, the Heavy Photon, which acquires a
weak coupling to electrically charged matter through kinetic mixing.
The HPS detector, a large acceptance forward spectrometer based on a dipole magnet, 
consists of a silicon tracker-vertexer, a lead-tungstate electromagnetic calorimeter, 
and a muon detector. HPS will search for the $e^{+}e^{-}$ or $\mu^{+}\mu^{-}$ decay
of the Heavy Photon produced in the interaction of high energy electrons with a high Z target, 
possibly with a displaced decay vertex. In this article, the description of the detector and 
its sensitivity are presented.
\end{abstract}

\section{The physics of the Heavy Photon Search experiment}
The nature of dark matter is one 
of the most important questions in particle physics. Recently,  an excess 
in the cosmic ray electron and positron spectra reported by the PAMELA experiment~\cite{pamela} 
has been confirmed and extended by other experiments~\cite{fermi}. 
One interesting possibility~\cite{nima} is that the signal can be explained 
by the existence of a new force, mediated by a massive, sub-GeV scale, 
U(1) gauge boson (the Heavy Photon or \Aprime) 
that couples very weakly to ordinary matter through "kinetic mixing"~\cite{holdom}.
TeV-scale dark matter could annihilate via an \Aprime\ boson which decay pre-dominantly into an 
$e^+e^-$ pair. This explanation is in accord with the dark 
matter relic abundance, the 
relatively large cross-section and the lack of excess in the baryon spectra for 
vector boson masses $<2\mathrm{m}_\mathrm{p}$. This weak coupling to the electric charge could be 
the only non-gravitational window into the existence of the hidden sector consisting of particles that do not 
couple to any of the known forces. Such hidden sectors are common in many new physics scenarios, 
see Ref.~\cite{Hewett:2012ns} for a recent review. 

Despite many existing constraints~\cite{darkforces}, 
there is a surprisingly large allowed parameter space to be examined by planned and proposed 
experiments. In the simplest scenarios, there are two main 
parameters that determine the characteristics of the \Aprime\ and thus the experimental search 
strategies: the kinetic mixing parameter $\epsilon \approx 10^{-12}-10^{-2}$ and the mass of the \Aprime. 
While a huge range of mixing parameters and masses are possible, it is natural that $\epsilon$ 
be around $10^{-3}$, and necessary that masses be around GeV if the positron excess is to be 
explained~\cite{Hewett:2012ns}. 
Experimentally, a very important  aspect is that in large parts of this parameter space the \Aprime\ can be long-lived 
with a lifetime proportional to $\sim \frac{1}{m_{\Aprime}^{2}\epsilon^2}$, and proper decay lengths varying from 
prompt to hundreds of meters~\cite{bible}. 
The Heavy Photon Search experiment (HPS) is a proposed fixed-target experiment~\cite{proposal_full}
specifically designed to discover an \Aprime\ with m$_{\Aprime}=10-1000$~MeV, produced through bremsstrahlung 
in a tungsten target and decaying into an $e^{+}e^{-}$ or $\mu^{+}\mu^{-}$ pair. 
In particular, the HPS experiment has sensitivity to the challenging region with small cross sections out of 
reach from collider experiments and where thick absorbers, as used in beam-dump experiments to 
reject backgrounds, are not allowed due to the relatively short \Aprime\ decay length ($<1$~m).  
This is accomplished by placing a compact silicon tracking and vertex detector in a magnetic field, 
immediately downstream (10~cm) of a thin ($\sim 0.25\%~X_{0} $) target. 
The \Aprime\ mass and decay vertex position is reconstructed from the measured momentum of its decay products.

\section{The HPS experiment}
The HPS experiment is proposed to run in Hall~B at the 
Thomas Jefferson National Accelerator Facility (JLab) with 
electron beam energies between 1.1 and 6.6~GeV and currents between 200-450~nA. 
\begin{figure}[t]
\centerline{\includegraphics[width=0.8\textwidth]{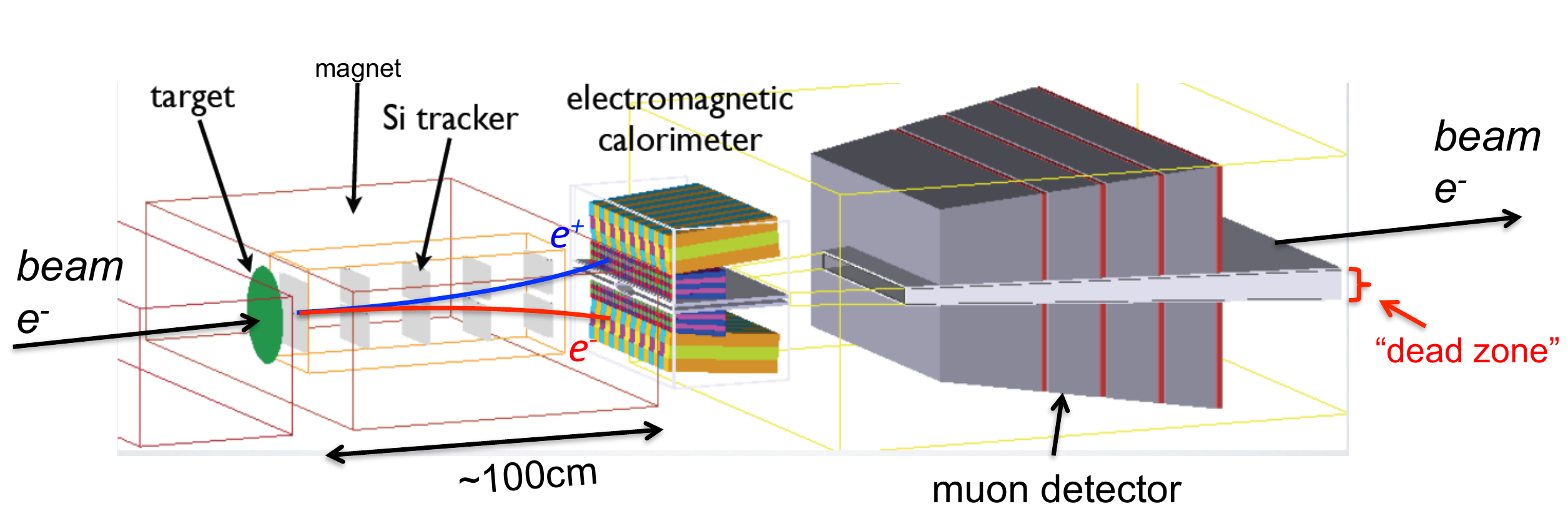}}
\caption{Conceptual design of the HPS detector.}
\label{fig:full_detector}
\end{figure}
The \Aprime\ will predominantly be produced within a few tens of mrad from the primary beam and carry most 
of the incoming electron's energy. The decay products for low $m_{\Aprime}$ thus appear at only a 
few tens of mrad requiring a detector with very forward acceptance, excellent event-time tagging 
capability to beat down backgrounds, and the ability to survive the intense amount of 
electromagnetic radiation generated in the target~\cite{bible}, especially close to the beam.
The silicon tracker, inside a magnetic dipole, will be placed above and below the beam plane, 
leaving a $\pm 15$~mrad "dead zone",  where the degraded beam 
can pass through the detector unobstructed, as shown in Fig.~\ref{fig:full_detector}. A lead-tungstate 
electromagnetic calorimeter with 250~MHz readout, placed above and below the beam plane 
downstream of the tracker, provides a trigger signal with 8~ns resolution. At m$_{\Aprime}$ above the 
di-muon threshold the calorimeter trigger is complemented by a scintillator based muon detector.
\section{The HPS silicon tracking and vertex detector}
\begin{figure}[]
\centerline{
\includegraphics[width=0.4\textwidth]{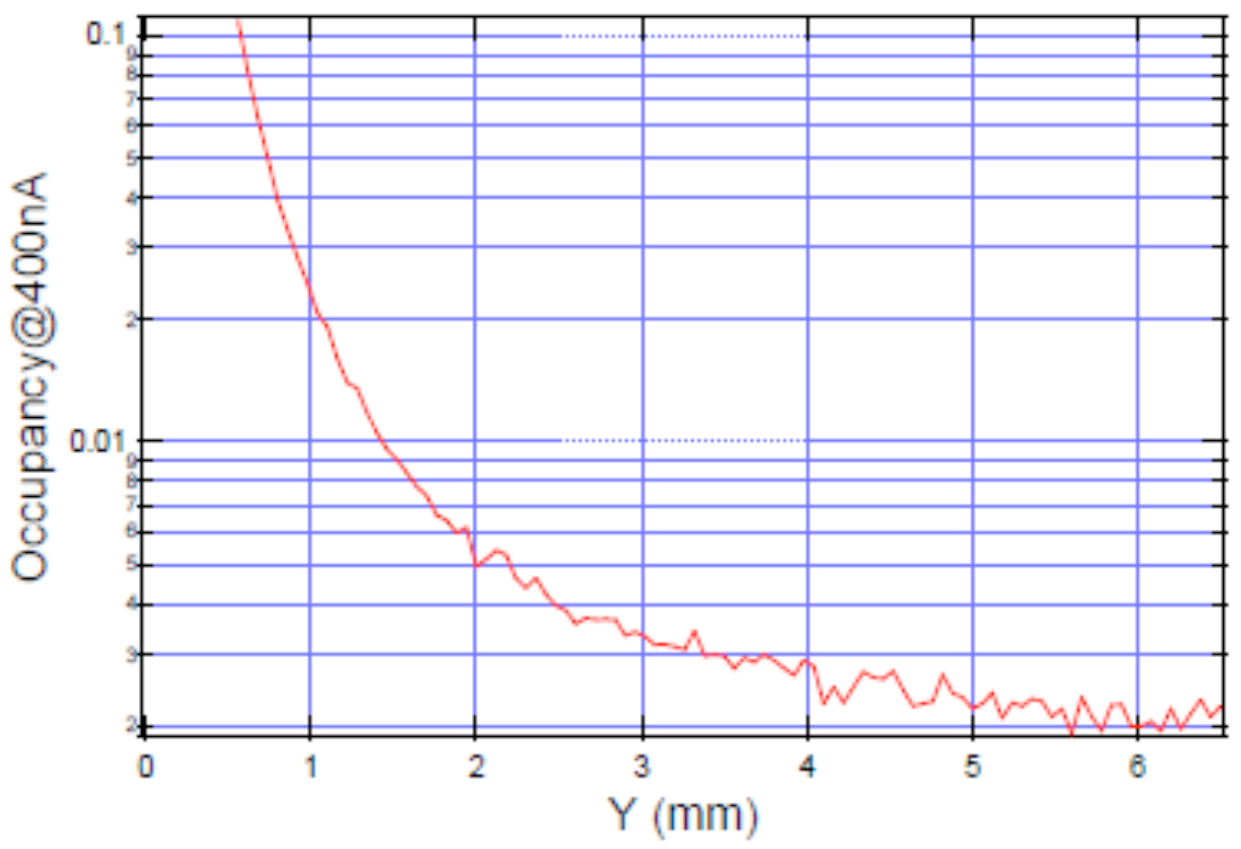}
\includegraphics[width=0.4\textwidth]{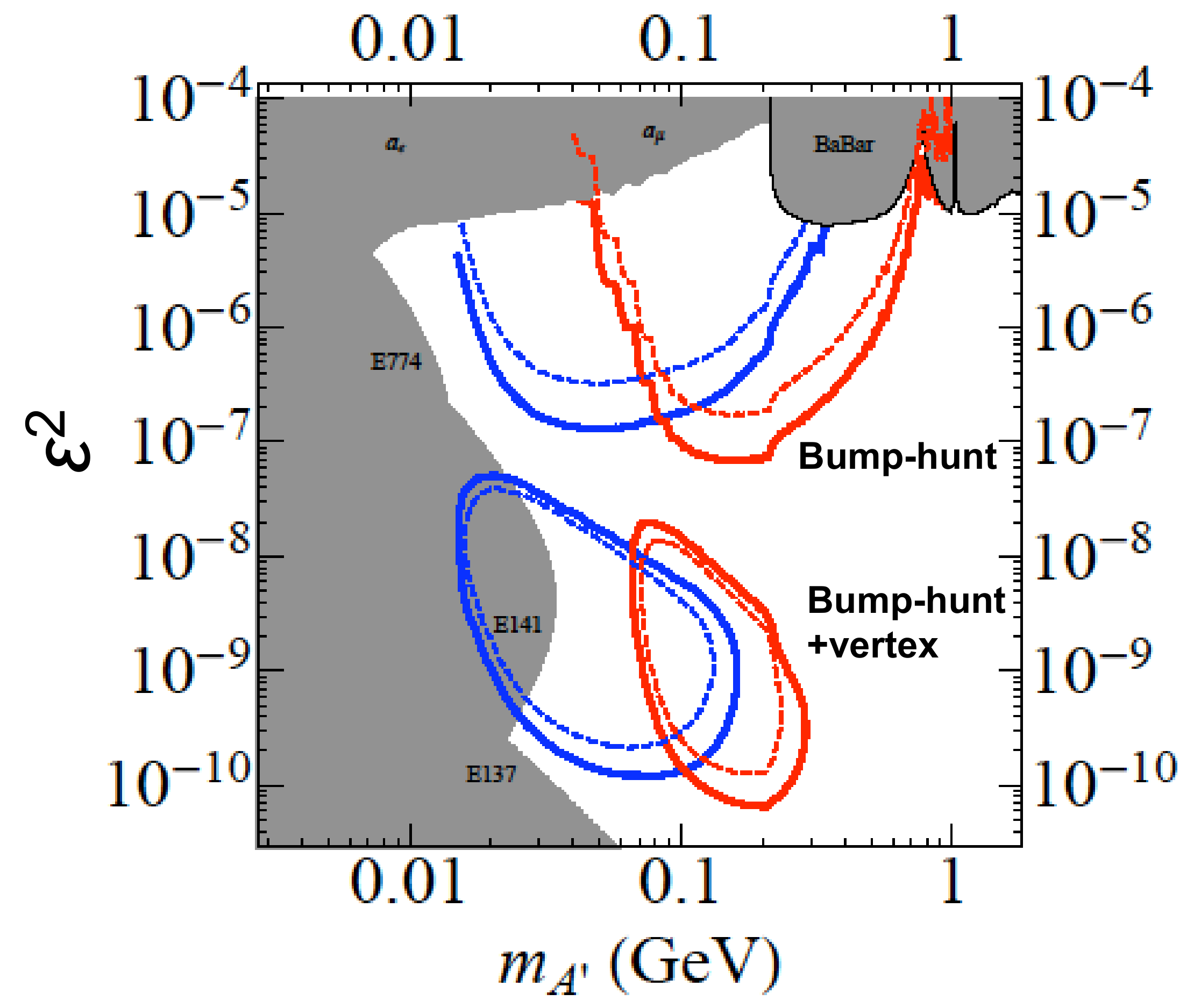}
}
\caption{\footnotesize Occupancy per strip ($60~\mu$m readout pitch) in tracking layer 1 for 
8~ns of beam at 400~nA (left). The reach for the HPS experiment at 2 (dashed) and 5 (solid) $\sigma$ significance (right).}
\label{fig:occupancy_and_reach}
\end{figure}
At beam energies necessary to achieve sensitivity to \Aprime\ in the most interesting mass range for HPS, 
multiple scattering dominates the measurement uncertainty, and in particular dictates the 
achievable vertex position resolution for any practical material budget. The main design guidelines are 
therefore to minimize the material budget in the tracking volume, and the distance to the beam in order 
to increase acceptance for low $m_{\Aprime}$, while keeping the occupancy under control.   
Furthermore, the whole tracker has to operate in vacuum to avoid secondary backgrounds from 
beam gas interactions, and have retractable tracking planes and easy access for sensor replacement to 
increase safety. 
Given the high hit density, the fast time response, and good resolution and radiation hardness needed; silicon microstrip sensors are the technology of choice for the tracker. Available pixel sensors had too large material 
budget. 
\begin{wraptable}{r}{0.39\textwidth}
\centerline{\begin{tabular}{|lcc|}
\hline
Layer $\rightarrow$& 1-3 & 4-6 \\ 
\hline
$z$ pos. (cm)  & 10-30 & 50-90  \\
Stereo angle  & $90^{\circ}$ & 50~mrad  \\
Bend res. ($\mu$m)  & $\approx 6$ & $\approx6$  \\
Stereo res. ($\mu$m)  & $\approx 6$ & $\approx130$  \\
\hline
\end{tabular}}
\caption{\footnotesize Main tracker parameters.}
\label{tab:svtparams}
\end{wraptable}
Each of the six tracking layers consists 
of a pair of Hamamatsu Photonics Corporation silicon microstrip sensors, with 90$^{\circ}$ or 100~mrad stereo angle, readily available at low cost 
from the cancelled D0 RunIIb upgrade~\cite{d0run2b}. These are $320~\mu$m thick, $p$+-on-$n$, 
AC coupled, polysilicon-biased sensors with 60 (30)~$\mu$m readout (sense) pitch and an overall 
size of 4x10~cm giving in total 67,480 channels. 
The optimized design, with the first layer placed only 10~cm downstream of the target to give 
excellent 3D vertexing performance, has a 15~mrad dead zone putting the active silicon only 
1.5~mm from the the center of the beam and hit densities locally reaching 4~MHz/cm$^2$ with 
occupancies kept $<1$\%, as shown in Fig.~\ref{fig:occupancy_and_reach}. 
To resolve overlapping hits in time and thus help to reject background and improve pattern recognition 
in the area closest to the beam, a 2~ns single hit time resolution is achieved by using the APV25 front-end readout 
ASIC initially developed for the CMS detector at CERN~\cite{apv25}. 
The APV25 chips, wire-bonded to 
the end of the sensor, are mounted on FR4 hybrid boards formed into multi-sensor modules with cooling 
and electrical services outside the tracking volume. The module support structure conducts heat from the 
sensor to integrated cooling under the hybrid board to remove $\sim1.7$~W of power.
The silicon operates at -8$^{\circ}$C to withstand high localized radiation doses up to 
$1 \times 10^{14}$~1~MeV neutron eq. fluence. Critical to reduce the multiple scattering uncertainty, a  
material budget of less than 0.7\%~$X_{0}$ per layer is obtained.
A vertical motion control system gives an adjustable distance to the beam plane for the top and 
bottom halves of the tracker, and mounting the whole structure on rails allows removal of the tracker from the 
vacuum chamber. 
With a precise 3D vertex resolution from the three 
most upstream layers with large stereo angles, and using the $\sim 20~\mu$m wide beam spot as a constraint, 
full track reconstruction simulation show a $\sim 10^7$ rejection of prompt backgrounds. 
This ensures good sensitivity for \Aprime\ decay lengths larger than 1~cm. 
Sensitivity to prompt \Aprime\ decays in a bump-hunt search is achieved 
by the best possible invariant mass resolution with a $\approx 6~\mu$m bend plane resolution in each 
tracking layer. Figure~\ref{fig:occupancy_and_reach} shows the expected sensitivity of the HPS experiment.

\section{The HPS test apparatus}
\begin{wrapfigure}{r}{0.4\textwidth}
\centerline{\includegraphics[width=0.4\textwidth]{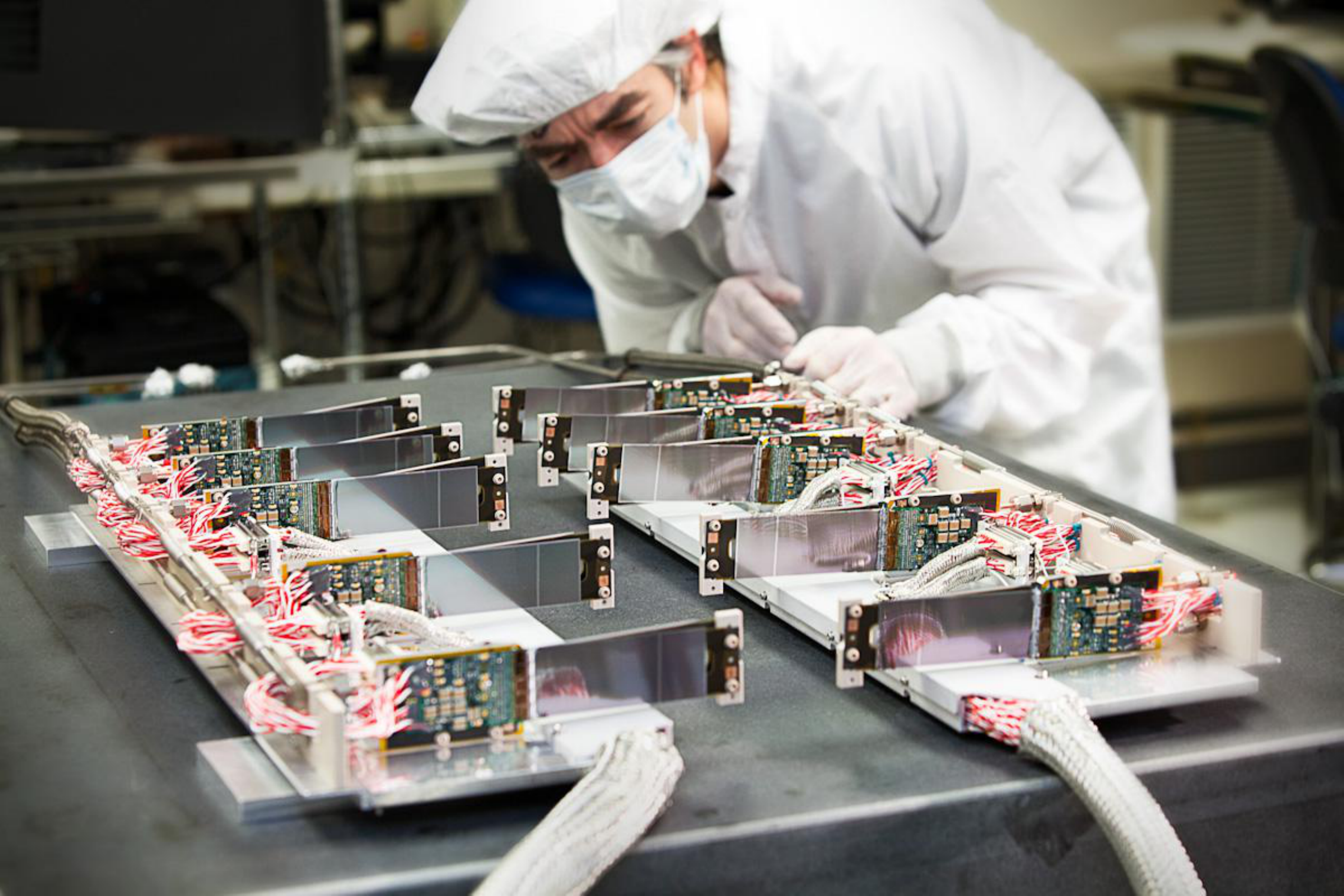}}
\caption{\footnotesize The two tracker halves of the HPS test tracker before integration.}
\label{fig:testsvt}
\end{wrapfigure}
The HPS Test run was proposed~\cite{proposal_testrun} to DOE early 2011 as a first stage of the HPS 
experiment 
and built based on the same design principles. The main difference, except the exclusion of the muon 
detector, is the smaller ECal and tracker acceptance with five layers and smaller stereo angles for 
simplicity and space constraints. 
The HPS test apparatus was installed and commissioned in April 2012, and while no dedicated 
electron beam was provided, it was able to collect valuable dedicated photon beam data.
Preliminary results include the verification of the modeling of the multiple Coulomb scattering, 
crucial for predicting occupancies for HPS, integration of the DAQ systems and full testing of the trigger system.

\section{Status and outlook}
After the test run, we have revisited the design of the full experiment, achieving nearly the same physics 
performance by building upon the strengths of the test run apparatus and mitigating its known weaknesses.   
This includes re-use of fundamental elements such as the ECal, trigger and silicon sensors with 
small improvements, increasing acceptance, and a 6th tracking layer. With this new proposal, HPS 
recycles the major beam line components and aims to be ready for data taking when first beam returns to 
Hall~B after the 12~GeV upgrade.


\begin{footnotesize}

\end{footnotesize}


\end{document}